\documentclass[twocolumn,showpacs,amsfonts,aps,prl,floatfix,%
groupedaddress]{revtex4} 
\usepackage{amsmath}
\usepackage{bm}
\usepackage{graphicx}

\newcommand{\be}[1]{\begin{equation}\label{#1}}
\newcommand{\ee}{\end{equation}}

\begin{document}

%%%%%%%%%%%%%%%%%%%%%%%%Front Matter%%%%%%%%%%%%%%%%%%%%%%%%%%%%%%%%%%
%%%%%%%%%%%%%%%%%%%%%%%%%%%%%%%%%%%%%%%%%%%%%%%%%%%%%%%%%%%%%%%%%%%%%%
\title{Multiplicity distribution and source deformation in full-overlap 
U+U collisions}
\date{\today}

\author{Anthony Kuhlman}
\author{Ulrich Heinz}
\email[Correspond to\ ]{heinz@mps.ohio-state.edu}
\affiliation{Department of Physics, The Ohio State University,
% 191 West Woodruff Avenue, 
  Columbus, OH 43210, USA}

\begin{abstract}
We present a full Monte Carlo simulation of the multiplicity and 
eccentricity distributions in U+U collisions at $\sqrt{s}=200\,A$\,GeV. 
While unavoidable trigger inefficiencies in selecting full-overlap U+U 
collisions cause significant modifications of the multiplicity distribution 
shown in \cite{Heinz:2005}, a selection of source eccentricities by cutting
the multiplicity distribution is still possible.
\end{abstract}


\maketitle
%%%%%%%%%%%%%%%%%%%%%%%%%%%%%%%%%%%%%%%%%%%%%%%%%%%%%%%%%%%%%%%%%%%%%%

In \cite{Heinz:2005} we advocated the use of full-overlap collisions
between deformed uranium nuclei to probe open questions at the 
Relativistic Heavy Ion Collider (RHIC). Specifically, such collisions 
can be used to test the hydrodynamic behavior of elliptic flow to much 
higher energy densities than currently possible with non-central Au+Au 
collisions, and the large and strongly deformed reaction zones produced 
in such collisions will allow for a detailed examination of the path 
length dependence of the energy lost by a fast parton as it travels 
through the plasma created in the collision.

The calculations presented in \cite{Heinz:2005} were based on the 
assumption that full-overlap U+U collisions can be efficiently
triggered on by using the zero degree calorimeters (ZDCs) of
the RHIC experiments to discriminate against collisions with 
spectator nucleons flying down the beam pipe. By cutting the
multiplicity distribution for the thus selected full-overlap 
collisions one can further select subevent classes with different
spatial deformations of the created fireballs. In this short note
we explore these assumptions in more quantitative detail, using a 
Monte Carlo simulation of the distributions of spectator nucleons
and charged particle multiplicity for U+U collisions for arbitrary
impact parameter and relative orientation between the two deformed 
uranium nuclei. 

Our calculations are based on a Glauber model parametrization of the
initial entropy production in these collisions, with standard 
Woods-Saxon form for the density distributions of the colliding 
nuclei (see \cite{KSH00} for details). [Possible modifications
arising from Color Glass Condensate (CGC) \cite{CGC} initial 
conditions \cite{Hirano:2004rs} will be shortly discussed at the 
end \cite{fn2}.] The initial entropy density in the transverse plane 
at $z{\,=\,}0$ (with $z$ denoting the beam direction) is determined 
by a combination of terms proportional to the wounded nucleon 
($n_{\rm wn}$) and binary collision ($n_{\rm bc}$) densities:
\be{eq:entropy_density}
   s(\bm{r}_\perp;\Phi) = \kappa_s\, 
   \left[\alpha\, n_{\rm wn}(\bm{r}_\perp;\Phi) 
         + (1{-}\alpha)\, n_{\rm bc}(\bm{r}_\perp;\Phi)\right] .
\ee
Here $\Phi$ is the angle between the beam direction and symmetry axis 
of one of the two U nuclei; in full-overlap collisions the symmetry 
axis of the other U nucleus lies in the same plane and forms an angle 
of $\Phi$ or $\pi{-}\Phi$ with the beam axis. The normalization constant 
$\kappa_s$ in \eqref{eq:entropy_density} is adjusted to reproduce the
charged particle multiplicity density $dN_{\rm ch}/dy$ measured at 
midrapidity in central 200\,$A$\,GeV Au+Au collisions at RHIC 
\cite{PHOBOSv2}, assuming proportionality of $dN_{\rm ch}/dy$ with 
%
%%%%%%%%%%%%%%%%%%%%%%%%%% Figure 1 %%%%%%%%%%%%%%%%%%%%%%%%%%%%%%%%%%
\begin{figure}
\includegraphics[width = \linewidth,clip]{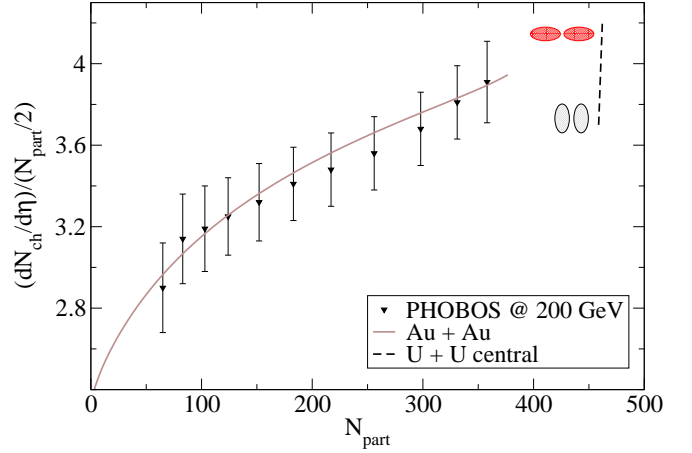}
\caption{Multiplicity per participant pair versus the number of
  participants $N_{\rm part}=\int d^2r_\perp\,n_{\rm wn}(\bm{r}_\perp)$ 
  for 200\,$A$\,GeV Au+Au and U+U collisions. Solid line: Glauber model 
  fit to the Au+Au data from PHOBOS \cite{PHOBOSv2}. Dashed line: 
  prediction for full-overlap U+U collisions.}
\label{F1}
\end{figure}
%%%%%%%%%%%%%%%%%%%%%%%%%%%%%%%%%%%%%%%%%%%%%%%%%%%%%%%%%%%%%%%%%%%%%%
%
the total entropy produced in the transverse plane. The wounded nucleon 
scaling fraction is tuned to $\alpha=0.75$ \cite{QGP3} to reproduce the
centrality dependence of $dN_{\rm ch}/dy$ (see Figure~\ref{F1}). 
After fitting $\kappa_s$ and $\alpha$ to the Au+Au data, we use the 
same parameters to predict the multiplicities for U+U collisions at 
the same $\sqrt{s}$. The results for full-overlap U+U collisions are
shown in Figure~\ref{F1} by the dashed line; low (high) multipli\-cities 
correspond to side-on-side (tip-on-tip) collisions as indicated, due
to their smaller (larger) binary collision contribution. 
%Side-on-side
%U+U collisions have a slightly smaller number of wounded nucleons 
%$N_{\rm part}$ than edge-on-edge collisions, due to their somewhat 
%larger surface contribution.

To determine the multiplicity distribution, we introduce Gaussian
event-by-event fluctuations of the multiplicity $n\equiv dN_{\rm ch}/dy$
via \cite{Kharzeev:2000ph}
\be{eq:prob_density}
  \frac{dP}{dn\ d\Phi} = A\ \exp\left\{-\frac{(n - \bar{n}(\Phi))^2}{2a
  \bar{n}(\Phi)}\right\},
\ee
where $\bar n(\Phi)$ is the average charged particle multiplicity
computed from Eq.~\eqref{eq:entropy_density} in a U+U collision with
orientation angle $\Phi$, and a width of $a=0.6$ has been shown to 
yield good agreement with PHOBOS data \cite{Kharzeev:2000ph}. The 
multiplicity distribution is then obtained by integrating 
\eqref{eq:prob_density} over $\Phi$. The resulting distribution 
\cite{Heinz:2005} is shown by the gray line in Figure \ref{F3} below.
Its double hump structure results from the Jacobian $d\bar n/d\Phi$,
and its asymmetry is a consequence of the fluctuation width being 
proportional to the mean multiplicity $\bar n$. Note that the non-linear
dependence of the charged multiplicity on the number of participant 
(wounded) nucleons, arising from the binary collision component in our 
parametrization \eqref{eq:entropy_density}, leads to a $\sim 15\%$ 
variation of the charged particle multiplicity among full-overlap 
U+U collisions as the relative orientation of their symmetry axes is 
varied over the accessible range.  

The calculations of multiplicity and eccentricity distributions 
presented in \cite{Heinz:2005} rely on the assumption that, by
monitoring spectator neutrons in the backward and forward zero
degree calorimeters (ZDCs) of the RHIC experiments, full-overlap 
collisions can be perfectly distinguished from those collisions 
where the two nuclei are slightly misaligned. This is impossible 
in practice since even fully aligned collisions in general have a 
small number of spectator nucleons, arising from the dilute nuclear 
surface, and this number is larger for side-on-side than for tip-on-tip 
collisions (see Figure~\ref{F1}). Therefore, slightly misaligned 
tip-on-tip and fully aligned side-on-side collisions can have 
the same $N_{\rm part}$ and the same ZDC signal. To assess the 
contamination from collisions with imperfect overlap on the
multiplicity and eccentricity distributions requires a more
comprehensive study which includes non-central U+U collisions.
This is the point of this short note.

\begin{figure}[htb]
\begin{center}
\includegraphics[width=\linewidth,clip]{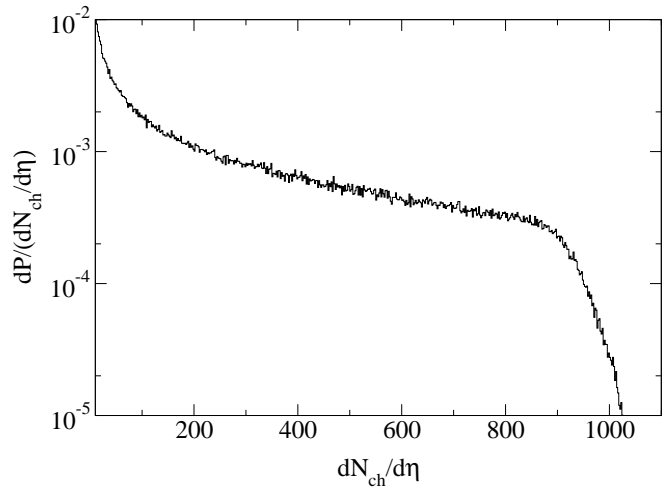}
\end{center}
\caption{Multiplicity distribution (normalized to unit total probability) 
  for $\sqrt{s}{\,=\,}200\,A$\,GeV U+U collisions. The distribution was
  generated from an impact parameter weighted sum of approximately 
  200,000 events with all possible orientations and impact parameters 
  between the two nuclei.}
\label{F2}
\end{figure}

A general U+U collision is parametrized by 5 parameters, the impact 
parameter $b$ and two Euler angles $\Omega=(\Phi,\beta)$ for each 
nucleus describing the orientation of its symmetry axis relative to 
the beam axis and impact parameter direction. Equation
\eqref{eq:entropy_density} for the initial entropy density must therefore
be generalized to
\begin{multline}\label{eq:entropy}
  s(\bm{r}_\perp; b, \Omega_1, \Omega_2) =
  \kappa_s\, \bigl[\alpha\,  n_{\rm wn}(\bm{r}_\perp; b, \Omega_1,\Omega_2)
  \\+ (1{-}\alpha)\,  n_{\rm bc}(\bm{r}_\perp; b,\Omega_1,\Omega_2)\bigr].
\end{multline}
The multiplicity distribution is then calculated from
\be{eq:prob2}
  \frac{dP}{dn} = A'\int b\,db\,d^2\Omega_1\,d^2\Omega_2 
  \exp\left\{-\frac{\left(n - \bar{n}(b,\Omega_1,\Omega_2)\right)^2}{2a
  \bar{n}(b,\Omega_1,\Omega_2)}\right\}.
\ee
Evaluating this 5-dimensional integral by Monte Carlo integration,
we obtain the multiplicity distribution shown in Figure~\ref{F2}.
Its right-most part contains the full-overlap collisions.

%%%%%%%%%%%%%%%%%%%%%%%%%% Figure 2 %%%%%%%%%%%%%%%%%%%%%%%%%%%%%%%%%%
%%%%%%%%%%%%%%%%%%%%%%%%%%%%%%%%%%%%%%%%%%%%%%%%%%%%%%%%%%%%%%%%%%%%%%

We can now try to select the latter from the overall event population 
by placing stringent cuts on the number of spectators 
($=2\times238-N_{\rm part}$). The distribution of the number of spectators
is shown in Figure~\ref{F5} below. In Figure~\ref{F3} we show the multiplicity 
distributions associated with the 5\% and 0.5\%, respectively, of events 
with the lowest spectator counts \cite{fn1}. It is immediately obvious 
that contamination from slightly misaligned collisions is sufficient to 
completely destroy the double-hump structure of the ideal full-overlap 
case, replacing it with a single peak. 
%
%%%%%%%%%%%%%%%%%%%%%%%%%% Figure 3 %%%%%%%%%%%%%%%%%%%%%%%%%%%%%%%%%%
\begin{figure}[hbt]
\begin{center}
\includegraphics[width=\linewidth,clip]{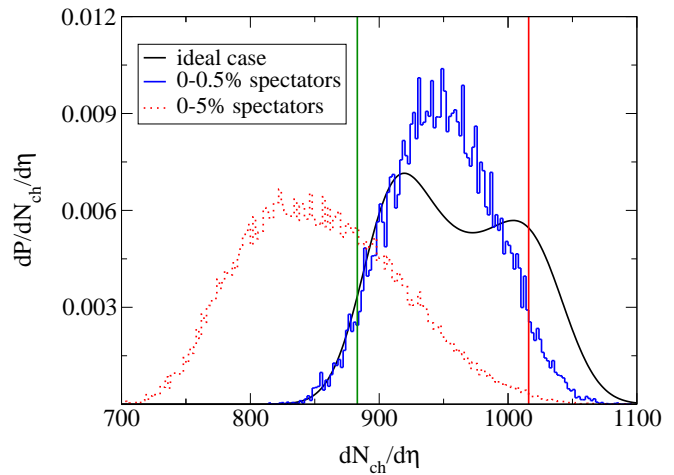}
\end{center}
\caption{Multiplicity distribution for the ideal case of only 
  full-overlap collisions (gray line), and for the 0.5\% and 5\%  
  U+U collision events with the lowest spectator counts 
  (solid blue and dotted red histograms, respectively ). All distributions 
  are normalized to unit total probability. The vertical lines indicate
  5\% and 95\% cuts on the blue multiplicity distribution for the 0.5\%
  spectators event class.}
\label{F3}
\end{figure}
%%%%%%%%%%%%%%%%%%%%%%%%%%%%%%%%%%%%%%%%%%%%%%%%%%%%%%%%%%%%%%%%%%%%%%
%
By selecting low-spectator events, we bias the sample towards events 
with $b \approx 0$, $\Phi_{1,2} \approx 0$, and the symmetry axes of 
the two nuclei approximately parallel. This suppresses the 
contribution from side-on-side configurations under the left peak of 
the idealized double-hump structure. At the same time, slightly
misaligned tip-on-tip collisions fill in the dip between the two humps 
from the idealized case \cite{fn2}. The result is a single-peaked 
multiplicity distribution whose center moves left (towards lower
multiplicities) as the cut on the number of spectator nucleons is
loosened.

Nevertheless, for sufficiently tight spectator cuts, we still expect 
the collision events corresponding to the left edge of the multiplicity 
distributions shown in Figure~\ref{F3} to have a larger contribution 
from strongly deformed side-on-side collisions than the events from the 
right edge (which will be mostly tip-on-tip collisions with small or 
zero source eccentricity). Following
%
%%%%%%%%%%%%%%%%%%%%%%%%%% Figure 4 %%%%%%%%%%%%%%%%%%%%%%%%%%%%%%%%%%
\begin{figure}[htb]
\begin{center}
\includegraphics[width=\linewidth,clip]{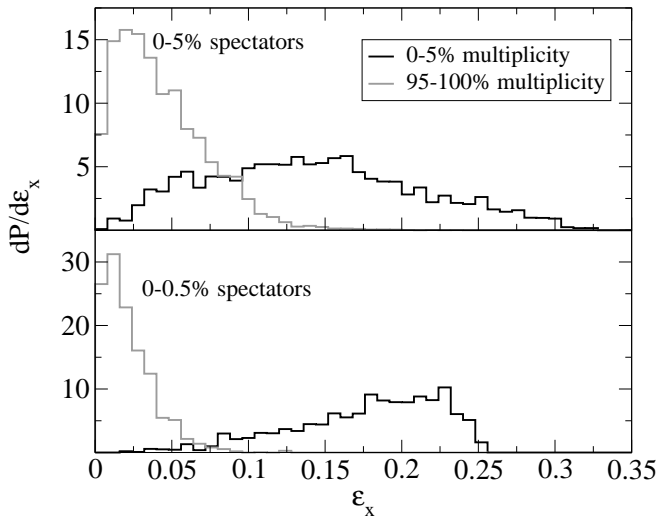}
\end{center}
\caption{Normalized eccentricity distributions, after making cuts on the
  multiplicity distributions for the two event classes shown in 
  Figure~\ref{F3} (top panel: 5\% spectator cut; bottom panel: 
  0.5\% spectator cut). Looser spectator cuts lead to broader 
  eccentricity distributions and less well-defined average 
  eccentricities.}
\label{F4}
\end{figure}
%%%%%%%%%%%%%%%%%%%%%%%%%%%%%%%%%%%%%%%%%%%%%%%%%%%%%%%%%%%%%%%%%%%%%%
%
our previous suggestion \cite{Heinz:2005} to select source eccentricities 
by cutting the multiplicity distribution of ``zero spectator'' collisions,
we perform such cuts on the more realistic distributions shown in 
Figure~\ref{F3}. Figure~\ref{F4} shows that it still possible in this 
way to select event classes with a given average source eccentricity: 
By taking the 0.5\% of events with the lowest spectator count from 
Figure~\ref{F3} (solid histogram) and cutting once more on the 5\% 
of events with the {\em lowest multiplicity}, we obtain the eccentricity 
distribution shown by the black histogram in the bottom panel of 
Figure~\ref{F4}. This event class has an average source deformation 
$\epsilon_x$ of about 18\%, corresponding to Au+Au collisions 
with impact parameters around 5.5\,fm. On the other hand, taking 
the same 0.5\% spectator cut and selecting the 5\% events with the 
{\em largest multiplicities} we obtain for the eccentricity 
distribution the gray histogram in the bottom Figure~\ref{F4}; this 
distribution peaks at $\epsilon_x=0$ and has a very small average 
spatial deformation.     

If one loosens the spectator cut to 5\% instead of 0.5\% (dotted histogram 
in Figure~\ref{F3}) and performs the same multiplicity selections
(5\% lowest or largest multiplicities, respectively), one obtains 
the eccentricity distributions shown in the top panel of 
Figure~\ref{F4}. Clearly, these distributions are much
broader than with the tighter spectator cut, and the average 
eccentricities shift down from 17.7\% to 14.2\% for the 
low-multiplicity selection and up from 2.2\% to 4.3\% for the 
high-multiplicity selection. Note that, since the looser spectator
cut allows for an increased contribution from non-zero impact
parameters, the eccentricity of the nuclear overlap region can
actually exceed the ${\approx\,}25\%$ ground state deformation
of the single-uranium density distribution projected on the 
transverse plane. This gives rise to the right tail of the black
histogram in the top panel of Figure~\ref{F4}. A typical event 
from this tail is shown in Figure~\ref{F5}. One sees that the 5\%
spectator cut allows for sizeable nonzero impact parameters and
numbers of spectator nucleons, and that very tight ZDC cuts are
required to ensure almost full overlap of the two uranium nuclei.  

%
%%%%%%%%%%%%%%%%%%%%%%%%%% Figure 5 %%%%%%%%%%%%%%%%%%%%%%%%%%%%%%%%%%
\begin{figure}[htb]
\begin{center}
\includegraphics[width=\linewidth,clip]{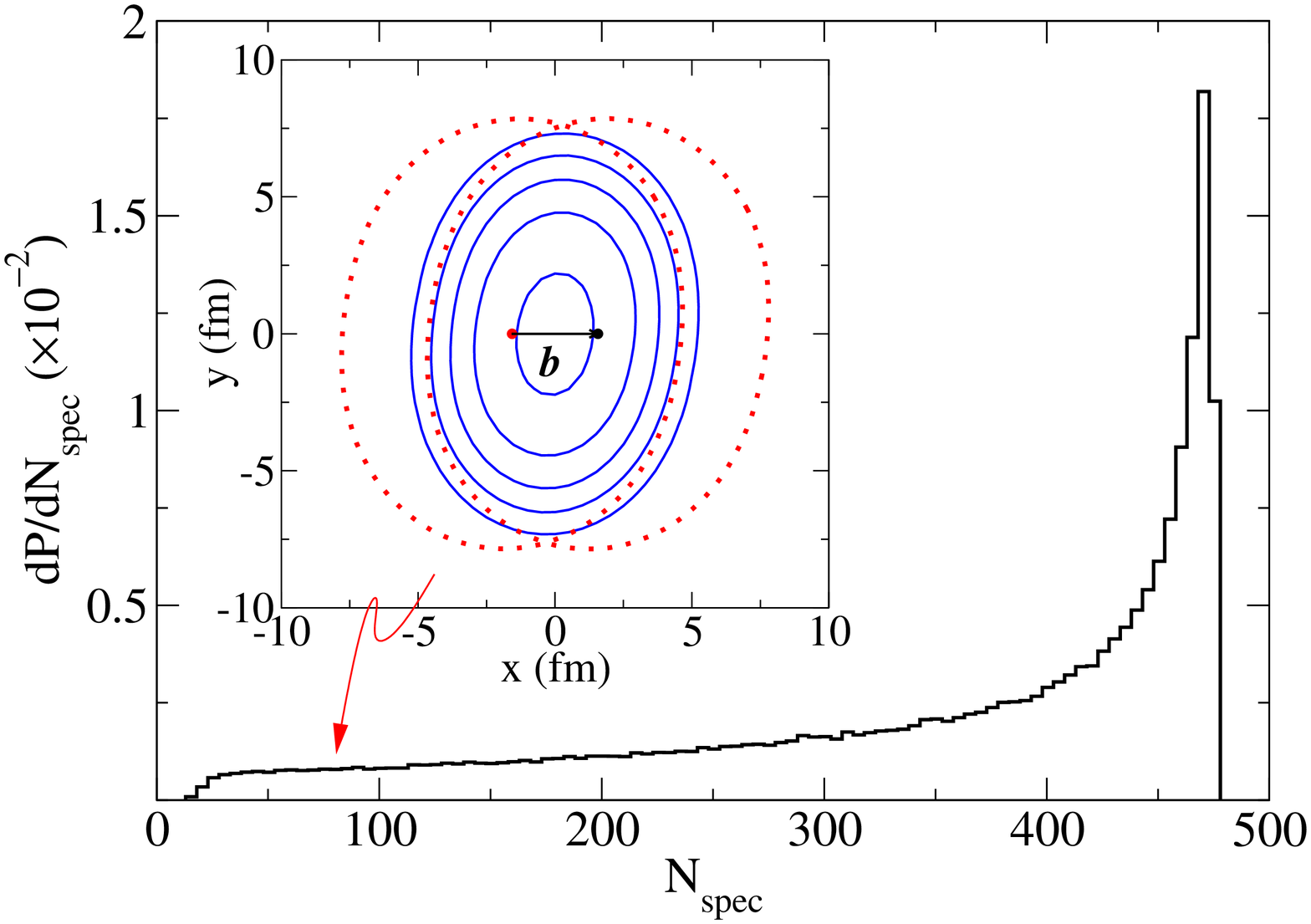}
\end{center}
\caption{Distribution of spectator nucleons for 200\,$A$\,GeV
  U+U collisions \cite{fn1}. The inset shows initial entropy density 
  contours ($s{\,=\,}20$, 40, 60, 80, 100\,fm$^{-3}$ from the outside 
  in) for a high-eccentricity event ($\epsilon_x{\,=\,}0.325$) which 
  is included in a ``5\% spectator cut'' event sample, as indicated by 
  the arrow.
  The impact parameter of this collision is $b{\,=\,}3.14\,$fm,
  and it produces $N_{\rm spec}{\,=\,}87$ spectator nucleons.}
\label{F5}
\end{figure}
%%%%%%%%%%%%%%%%%%%%%%%%%%%%%%%%%%%%%%%%%%%%%%%%%%%%%%%%%%%%%%%%%%%%%%
%

The detailed shapes of the eccentricity distributions shown in 
Figure~\ref{F4} are expected to depend somewhat on our parametrization 
(\ref{eq:entropy_density},\ref{eq:entropy}) of the initial
transverse density distribution of the produced matter. It was shown
in Ref.~\cite{Hirano:2004rs} that initial conditions motivated by
the Color Glass Condensate picture of low-$x$ gluon saturation in
large nuclei at high energies \cite{CGC} produce transverse density
distributions which fall off more steeply near the edge than the more
Gaussian-like distributions \cite{Kolb:2001qz} resulting from our
Eqs.~(\ref{eq:entropy_density},\ref{eq:entropy}). This might result 
in somewhat larger eccentricities and narrower eccentricity distributions
(i.e. better defined average eccentricities) than those shown in 
Figure~\ref{F4}. However, significantly higher statistics would 
likely be needed to clearly see such differences. 

These results show that the suggestions made in \cite{Heinz:2005} for
using U+U collisions to explore in more detail the ideal fluid
dynamic nature of elliptic flow and the path length dependence
of radiative parton energy loss are reasonably robust against trigger
inefficiencies, and that a meaningful U+U collision program at RHIC is, 
in fact, feasible \cite{fn3}. Simultaneous strict cuts on small numbers 
of spectator nucleons and on charged particle multiplicity are necessary
to select collisions which produce sources with well-defined and large
spatial deformation; the histograms shown in the bottom panel of 
Figure~\ref{F4} correspond to only 0.025\% of all U+U collisions taking 
place in the accelerator. The top panel in Figure~\ref{F4} shows
that it is possible to loosen these tight cuts somewhat, at the expense 
of reducing the average spatial source deformation and introducing 
larger event-by-event fluctuations as well as an increased sensitivity 
to details of the Glauber model used for relating the relative 
nuclear orientation to the observed spectator nucleon and charged 
hadron multiplicities. We leave a further discussion of such model 
uncertainties (see also \cite{fn2}) for later when U+U collisions 
become (hopefully) available. 

We thank M. Gyulassy for stimulating discussions and G. Fai for an 
enlightening question which led to the discovery of an error in 
Eq.~(\ref{eq:prob2}) in the originally submitted manuscript. This 
work was supported by the U.S. Department of Energy under contract 
DE-FG02-01ER41190.


\begin{thebibliography}{99}
\bibitem{fn1}
Our calculations neglect event-by-event fluctuations of the number of
spectators in collisions of given impact para\-meter $b$ and orientations
$\Omega_1,\,\Omega_2$. We leave this for a more detailed Monte Carlo 
simulation when U+U collisions become available.
\end{thebibliography}
\end{document}